%% file: stoffer_kl4.tex
\pdfoutput=1
\documentclass[epjCONF,final]{svjour}
\usepackage{graphics}
\usepackage[varg]{txfonts} 
\usepackage[latin1]{inputenc}
\usepackage[UKenglish]{babel}
\usepackage{graphicx}
\usepackage{color}

\usepackage{savesym}
\savesymbol{iint}
\savesymbol{iiint}
\savesymbol{iiiint}
\savesymbol{idotsint}
\usepackage{amsmath}

\usepackage{url}
\usepackage{hyperref}

\usepackage[absolute]{textpos}

\renewcommand{\Im}{\mathrm{Im}}

\newcommand{\<}{\langle}
\renewcommand{\>}{\rangle}

\renewcommand{\O}{\mathcal{O}}

\session-title{MESON2012 - 12$^{\text{\footnotesize th}}$ International Workshop on Meson Production, Properties and Interaction}
\begin{document}
\title{A Dispersive Treatment of $K_{\ell4}$~Decays}
\author{Gilberto Colangelo\inst{1} \and 
Emilie Passemar\inst{2} \and 
\underline{Peter Stoffer}\inst{1}\fnmsep\thanks{\email{stoffer@itp.unibe.ch}} }
\institute{Albert Einstein Center for Fundamental Physics, Institute for Theoretical Physics, University of Bern, Sidlerstrasse 5, CH-3012 Bern, Switzerland \and 
Theoretical Division, Los Alamos National Laboratory, Los Alamos, NM 87545, USA}
\abstract{
$K_{\ell4}$ are for several reasons an especially interesting decay channel of $K$ mesons:
$K_{\ell4}$ decays allow an accurate measurement of a combination of
$S$-wave $\pi\pi$ scattering lengths, one form factor of the decay is
connected to the chiral anomaly and the decay is the best source for the
determination of some low energy constants of ChPT.  We present a dispersive
approach to $K_{\ell4}$ decays, which takes rescattering effects
fully into account.  Some fits to NA48/2 and E865 measurements and results of the
matching to ChPT are shown.
} 
\maketitle
%

	\section{Motivation}
	\label{sec:Motivation}
	
	$K_{\ell4}$, the semileptonic decay of a kaon into two pions and a lepton-neutrino pair, plays a crucial role in the context of low energy hadron physics, because it provides almost unique information about some of the $\O(p^4)$ low energy constants (LECs) of Chiral Perturbation Theory (ChPT), the effective low energy theory of QCD. The physical region of $K_{\ell4}$ starts already at the $\pi\pi$ threshold, thus it happens at lower energies than e.g.~elastic $K\pi$ scattering. Since ChPT is an expansion in the masses and momenta, it is expected to converge better at lower energies. Therefore $K_{\ell4}$ is a particularly interesting process to study.
	
	Besides, as the hadronic final state contains two pions, $K_{\ell4}$ is also one of the best sources of information on the $\pi\pi$ scattering lengths $a_0^0$ and $a_0^2$ \cite{ScatteringLengths}.
	
	On the experimental side, we are confronted with impressive precision from high statistics measurements. During the last decade, the process has been measured in the E865 experiment at BNL \cite{E865} and in the NA48/2 experiment at CERN \cite{ScatteringLengths}. Very recently, the NA48/2 collaboration has published the results on the branching ratio and form factors of $K_{\ell4}$, based on more than a million events \cite{NA48}.
	
	Here, we present preliminary results of a new dispersive treatment of $K_{\ell4}$ decays. Dispersion relations are an interesting tool to treat low energy hadronic processes. They are based on the very general principles of analyticity and unitarity. The derivation of the dispersion relation applies a chiral power counting and is valid up to and including $\O(p^6)$. The dispersion relation is parametrised by five subtraction constants. As soon as these constants have been fixed, the energy dependence is fully determined by the dispersion relation. The presented method implements a summation of final state rescattering effects, thus we expect it to incorporate the most important contributions beyond $\O(p^6)$.
	
\setlength{\TPHorizModule}{1cm}
\setlength{\TPVertModule}{1cm}
\begin{textblock}{4}(13.75,3)
LA-UR-12-23703
\end{textblock}

	\newpage
	
	\section{Dispersion Relation for $K_{\ell4}$ Decays}
	\label{sec:DispersionRelation}
	
	\subsection{Matrix Element and Form Factors}
	\label{sec:MatrixElement}
	
	We consider the charged decay mode of $K_{\ell4}$:
	\begin{align}
		K^+(p) \rightarrow \pi^+(p_1) \pi^-(p_2) \ell^+(p_\ell) \nu_\ell(p_\nu) ,
	\end{align}
	where $\ell\in\{e,\mu\}$ is either an electron or a muon.

	After integrating out the $W$ boson, we end up with a Fermi type current-current interaction and the matrix element splits up into a leptonic times a hadronic part. The leptonic matrix element can be treated in a standard way. The hadronic matrix element exhibits the usual $V-A$ structure of weak interaction. Its Lorentz structure allows us to write the two contributions as
	\begin{align}
		\big\< \pi^+(p_1) \pi^-(p_2) \big| V_\mu(0) \big| K^+(p)\big\> &= -\frac{H}{M_K^3} \epsilon_{\mu\nu\rho\sigma} L^\nu P^\rho Q^\sigma , \\
		\big\< \pi^+(p_1) \pi^-(p_2) \big| A_\mu(0) \big| K^+(p) \big\> &= -i \frac{1}{M_K} \left( P_\mu F + Q_\mu G + L_\mu R \right) .
	\end{align}
	In the electron mode (up to now the only one where experimental data is available), mainly one specific linear combination of the form factors $F$ and $G$ is accessible:
	\begin{align}
		F_1(s,t,u) = X F(s,t,u) + (u-t) \frac{PL}{2X} G(s,t,u) ,
	\end{align}
	where $s$, $t$ and $u$ are the usual Mandelstam variables, $X = \frac{1}{2}\lambda^{1/2}(M_K^2, s, s_\ell)$, $PL = \frac{1}{2}(M_K^2 - s - s_\ell)$ and $s_\ell = (p_\ell + p_\nu)^2$. $\lambda(a,b,c) = a^2 + b^2 + c^2 - 2(ab + bc + ca)$ is the K\"all\'en triangle function.

	\subsection{Decomposition of the Form Factor}
	\label{sec:Decomposition}	
	
	The form factor $F_1$ has the following analytic properties:
	\begin{itemize}
		\item There is a right-hand branch cut in the complex $s$-plane, starting at the $\pi\pi$-threshold.
		\item Analogously, in the $t$- and $u$-channel, right-hand cuts start at the $K\pi$-threshold.
	\end{itemize}
	Due to crossing, the right-hand cuts in the $t$- and $u$-channel show up in the $s$-channel for negative values of $s$. The situation is analogous for the other channels.
	
	Based on fixed-$t$ and fixed-$u$ dispersion relations for the form factor, we can derive its decomposition into functions of a single variable. Such a decomposition has first been worked out for the $\pi\pi$ scattering amplitude \cite{Stern1993} and later for $K\pi$ scattering \cite{Ananthanarayan2001}.
	
	The basic idea is to define functions that only contain the right-hand cut of each partial wave and to split up in this way all the discontinuities of the form factor. For instance, in the case of the $s$-channel $S$-wave $f_0$, this function looks as follows:
	 \begin{align}
	 	\label{eqn:M0DispRel}
		M_0(s) := P(s) + \frac{s^4}{\pi} \int_{4M_\pi^2}^{\Lambda^2} \frac{\Im f_0(s^\prime)}{(s^\prime - s - i\epsilon) {s^\prime}^4} ds^\prime ,
	\end{align}
	where $P(s)$ is a subtraction polynomial. After defining similar functions that take care of the right-hand cuts of $f_1$ and the $S$- and $P$-waves in the crossed channels, all the discontinuities are divided into functions of a single variable.
	
	This procedure, also known as `reconstruction theorem', neglects on the one hand the imaginary parts of $D$- and higher waves, on the other hand the high energy tails of the dispersion integrals from $\Lambda^2$ to $\infty$. Both effects are of $\O(p^8)$ in the chiral counting. In order to simplify substantially the dispersion relation, we also neglect at the present preliminary stage the dependence on $s_\ell$, which is experimentally small. (Otherwise, we would need to consider a coupled system for the form factors $F$ and $G$.)
	
	Respecting isospin properties, we obtain the following decomposition of the form factor:
	\begin{align}
		\label{eqn:Decomposition}
		\begin{split}
			F_1(s,t,u) &= M_0(s) + \frac{2}{3} N_0(t) + \frac{1}{3} R_0(t) + R_0(u) + (u-t) M_1(s) - \frac{2}{3} \Big[ t (u-s) - \Delta_{K\pi}\Delta_{\ell\pi} \Big] N_1(t) ,
		\end{split}
	\end{align}
	where  $\Delta_{K\pi} = M_K^2 - M_\pi^2$ and $\Delta_{\ell\pi} = s_\ell - M_\pi^2$.
	
	\subsection{Integral Equations}
	\label{sec:IntegralEquations}
	
	One should not try to solve directly the dispersion relation~(\ref{eqn:M0DispRel}) since it may not uniquely determine the solution of the problem \cite{Leutwyler1996}. Noting that each of the functions $M_0$, $\ldots$ satisfies an Omn\`es equation, we apply the solution to the inhomogeneous Omn\`es problem:
	\begin{align}
		M_0(s) &= \Omega_0^0(s) \Bigg\{ \tilde P(s) + \frac{s^3}{\pi} \int_{4M_\pi^2}^{\Lambda^2} \frac{\hat M_0(s^\prime) \sin \delta_0^0(s^\prime)}{|\Omega_0^0(s^\prime)| (s^\prime - s - i \epsilon){s^\prime}^3} ds^\prime \Bigg\} ,
	\end{align}
	with a new subtraction polynomial $\tilde P(s)$ and the Omn\`es function
	\begin{align}
		\Omega_0^0(s) := \exp\left\{ \frac{s}{\pi} \int_{4M_\pi^2}^\infty \frac{\delta_0^0(s^\prime)}{s^\prime(s^\prime-s-i\epsilon)} \, ds^\prime \right\} .
	\end{align}
	Similar relations hold for the other functions. Due to Watson's final state theorem, we can identify the phases $\delta_l^I$ ($l$ -- angular momentum, $I$ -- isospin) with the phase shifts of the elastic scattering. As an input to our equations, we therefore need the following phase shifts:
	\begin{itemize}
		\item $\delta_0^0$, $\delta_1^1$: elastic $\pi\pi$ scattering  \cite{Colangelo2011}
		\item $\delta_0^{1/2}$, $\delta_1^{1/2}$, $\delta_0^{3/2}$: elastic $K\pi$ scattering  \cite{Moussallam2004,Boito2010}
	\end{itemize}
	
	The inhomogeneities in the Omn\`es problem are given by the differences of the functions $M_0$,~$\ldots$ and the corresponding partial wave, e.g.~$\hat M_0(s) = f_0(s) - M_0(s)$. These `hat functions' contain the left-hand cut of the partial wave and we compute them by projecting out the partial wave of the decomposed form factor~(\ref{eqn:Decomposition}). E.g.~$\hat M_0(s)$ is then given as angular averages of $N_0$, $N_1$, etc.
	
	We can now solve the dispersion relation for the form factor. We have parametrised the problem by the constants appearing in the subtraction polynomials (in total, only five independent subtraction constants are needed due to an ambiguity in the decomposition) and we use the elastic scattering phase shifts as inputs. The energy dependence is then fully determined by the dispersion relation.
	
	We face a set of coupled integral equations: The functions $M_0(s)$, $M_1(s)$, $\dots$ are given as dispersive integrals involving the hat functions $\hat M_0(s)$, $\hat M_1(s)$, $\ldots$, whereas the hat functions are themselves defined as angular integrals over $M_0(s)$, $M_1(s)$, etc. This system can be solved by iteration. The problem is linear in the subtraction constants that have to be determined by a fit to data.

	\section{Preliminary Results}
	\label{sec:Results}

	\subsection{Fit to Data}
	\label{sec:FitToData}
	
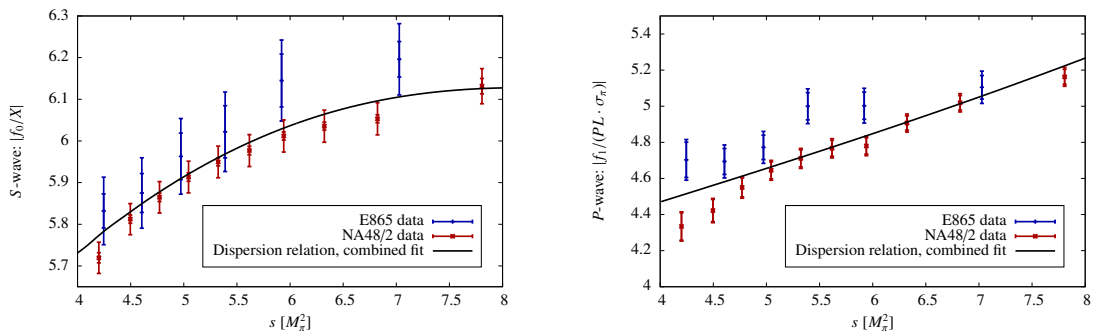
\begin{figure}[ht]
		\centering
		\large
		\scalebox{0.5}{
			\input{plots/comparison_DR_data_swave.tex}
		}
		\large
		\hspace{1cm}
		\scalebox{0.5}{
			\input{plots/comparison_DR_data_pwave.tex}
		}
		\caption{Fit of the $S$- and $P$-wave to both E865 \cite{E865} and NA48/2 \cite{NA48} data. The error bars indicate separately the uncorrelated part of the uncertainties and the total errors (containing also the uncertainty from the normalisation).}
		\label{fig:FitToSWave}
	\end{figure}

	We perform a fit of the dispersion relation to both, the E865 \cite{E865} and NA48/2 data sets \cite{NA48}. The $S$-wave dominantly determines the three subtraction constants in $M_0$, the $P$-wave the two in $M_1$. Figure~\ref{fig:FitToSWave} shows the result of the combined fit for the two partial waves. The $\chi^2/\mathrm{dof}$ of this fit is 1.73. This rather large value is a consequence of the small uncertainties of the relative form factors in the NA48/2 data.

	\subsection{Matching to ChPT}
	\label{sec:MatchingChPT}
	
	After having fixed the subtraction constants by fitting the data, we can match our dispersive representation to the ChPT result. In this matching procedure, we determine the LECs $L_1^r$, $L_2^r$ and $L_3^r$.
	
	The dispersion relation allows us to choose a convenient matching point that may lie outside the physical region. We choose the point $s=t-u=0$ below the threshold, where ChPT should even converge better. In table~\ref{tab:PreliminaryLECs}, we show preliminary values of the low energy constants, resulting from a matching to the $\O(p^4)$ ChPT result. For comparison, we also quote the values of the global fit by Bijnens and Jemos \cite{Bijnens2011} that takes as input the $K_{\ell4}$ data from NA48/2 \cite{ScatteringLengths}.
	\begin{table}[ht]
		\centering
		\caption{Preliminary results for the LECs ($\mu = 770$~MeV).}
		\label{tab:PreliminaryLECs}
		\begin{tabular}{lccc}
			\hline\noalign{\smallskip}
			 & $10^3 L_1^r$ & $10^3 L_2^r$ & $10^3 L_3^r$ \\
			\noalign{\smallskip}\hline\noalign{\smallskip}
			Dispersive treatment, fit to E865 \cite{E865} & $0.42 \pm 0.41$ & $0.41 \pm 0.34$ & $-2.19 \pm 1.41$ \\
			Dispersive treatment, fit to NA48/2 \cite{NA48} & $0.60 \pm 0.29$ & $0.63 \pm 0.28$ & $-3.16 \pm 1.19$ \\
			Bijnens, Jemos, `fit All' \cite{Bijnens2011} & $0.88 \pm 0.09$ & $0.61 \pm 0.20$ & $-3.04 \pm 0.43$ \\
			\noalign{\smallskip}\hline
		\end{tabular}
	\end{table}

	We find that our fit to NA48/2 agrees with this global fit. A large part of our uncertainties stems from the $\pi\pi$ scattering phase shifts and contains at present a conservative estimate of the systematics.
	
	Our preliminary values are subject to change as we are going to perform a matching to $\O(p^6)$ ChPT and include the neglected $s_\ell$ dependence as well as isospin breaking corrections \cite{InPreparation}.
	
	\subsection{Summary}
	\label{sec:Summary}
	
	We have presented a dispersive representation of $K_{\ell4}$ decays that provides a model independent parametrisation valid up to and including $\O(p^6)$. It includes a full summation of final state rescattering effects. It is parametrised by subtraction constants that we fix by fitting experimental data. The matching to ChPT can be performed below the physical threshold, where ChPT should converge better. Hence, we expect to find more reliable values for the LECs than with a pure ChPT treatment.

	\section*{Acknowledgements}
	\label{sec:Acknowledgements}
	
	The speaker (PS) thanks the local conference committee of the Jagiellonian University for the perfect organisation of the MESON2012 conference. We are very grateful to B.~Bloch-Devaux for her support. We further thank J.~Bijnens, J.~Gasser, B.~Kubis, S.~Lanz, H.~Leutwyler, S.~Pislak and P.~Tru\"ol for useful discussions. The Albert Einstein Center for Fundamental Physics is supported by the `Innovations- und Kooperationsprojekt C--13' of the `Schweizerische Universit\"atskonferenz SUK/CRUS'. This work was supported in part by the Swiss National Science Foundation.

\end{document}

%% file: plots/comparison_DR_data_swave.tex
\begingroup
  \makeatletter
  \providecommand\color[2][]{%
    \GenericError{(gnuplot) \space\space\space\@spaces}{%
      Package color not loaded in conjunction with
      terminal option `colourtext'%
    }{See the gnuplot documentation for explanation.%
    }{Either use 'blacktext' in gnuplot or load the package
      color.sty in LaTeX.}%
    \renewcommand\color[2][]{}%
  }%
  \providecommand\includegraphics[2][]{%
    \GenericError{(gnuplot) \space\space\space\@spaces}{%
      Package graphicx or graphics not loaded%
    }{See the gnuplot documentation for explanation.%
    }{The gnuplot epslatex terminal needs graphicx.sty or graphics.sty.}%
    \renewcommand\includegraphics[2][]{}%
  }%
  \providecommand\rotatebox[2]{#2}%
  \@ifundefined{ifGPcolor}{%
    \newif\ifGPcolor
    \GPcolortrue
  }{}%
  \@ifundefined{ifGPblacktext}{%
    \newif\ifGPblacktext
    \GPblacktexttrue
  }{}%
  \let\gplgaddtomacro\g@addto@macro
  \gdef\gplbacktext{}%
  \gdef\gplfronttext{}%
  \makeatother
  \ifGPblacktext
    \def\colorrgb#1{}%
    \def\colorgray#1{}%
  \else
    \ifGPcolor
      \def\colorrgb#1{\color[rgb]{#1}}%
      \def\colorgray#1{\color[gray]{#1}}%
      \expandafter\def\csname LTw\endcsname{\color{white}}%
      \expandafter\def\csname LTb\endcsname{\color{black}}%
      \expandafter\def\csname LTa\endcsname{\color{black}}%
      \expandafter\def\csname LT0\endcsname{\color[rgb]{1,0,0}}%
      \expandafter\def\csname LT1\endcsname{\color[rgb]{0,1,0}}%
      \expandafter\def\csname LT2\endcsname{\color[rgb]{0,0,1}}%
      \expandafter\def\csname LT3\endcsname{\color[rgb]{1,0,1}}%
      \expandafter\def\csname LT4\endcsname{\color[rgb]{0,1,1}}%
      \expandafter\def\csname LT5\endcsname{\color[rgb]{1,1,0}}%
      \expandafter\def\csname LT6\endcsname{\color[rgb]{0,0,0}}%
      \expandafter\def\csname LT7\endcsname{\color[rgb]{1,0.3,0}}%
      \expandafter\def\csname LT8\endcsname{\color[rgb]{0.5,0.5,0.5}}%
    \else
      \def\colorrgb#1{\color{black}}%
      \def\colorgray#1{\color[gray]{#1}}%
      \expandafter\def\csname LTw\endcsname{\color{white}}%
      \expandafter\def\csname LTb\endcsname{\color{black}}%
      \expandafter\def\csname LTa\endcsname{\color{black}}%
      \expandafter\def\csname LT0\endcsname{\color{black}}%
      \expandafter\def\csname LT1\endcsname{\color{black}}%
      \expandafter\def\csname LT2\endcsname{\color{black}}%
      \expandafter\def\csname LT3\endcsname{\color{black}}%
      \expandafter\def\csname LT4\endcsname{\color{black}}%
      \expandafter\def\csname LT5\endcsname{\color{black}}%
      \expandafter\def\csname LT6\endcsname{\color{black}}%
      \expandafter\def\csname LT7\endcsname{\color{black}}%
      \expandafter\def\csname LT8\endcsname{\color{black}}%
    \fi
  \fi
  \setlength{\unitlength}{0.0500bp}%
  \begin{picture}(7200.00,5040.00)%
    \gplgaddtomacro\gplbacktext{%
      \csname LTb\endcsname%
      \put(660,1017){\makebox(0,0)[r]{\strut{} 5.7}}%
      \put(660,1643){\makebox(0,0)[r]{\strut{} 5.8}}%
      \put(660,2270){\makebox(0,0)[r]{\strut{} 5.9}}%
      \put(660,2896){\makebox(0,0)[r]{\strut{} 6}}%
      \put(660,3522){\makebox(0,0)[r]{\strut{} 6.1}}%
      \put(660,4149){\makebox(0,0)[r]{\strut{} 6.2}}%
      \put(660,4775){\makebox(0,0)[r]{\strut{} 6.3}}%
      \put(792,484){\makebox(0,0){\strut{} 4}}%
      \put(1585,484){\makebox(0,0){\strut{} 4.5}}%
      \put(2377,484){\makebox(0,0){\strut{} 5}}%
      \put(3170,484){\makebox(0,0){\strut{} 5.5}}%
      \put(3963,484){\makebox(0,0){\strut{} 6}}%
      \put(4755,484){\makebox(0,0){\strut{} 6.5}}%
      \put(5548,484){\makebox(0,0){\strut{} 7}}%
      \put(6340,484){\makebox(0,0){\strut{} 7.5}}%
      \put(7133,484){\makebox(0,0){\strut{} 8}}%
      \put(-110,2739){\rotatebox{-270}{\makebox(0,0){\strut{}$S$-wave: $|f_0/X|$}}}%
      \put(3962,154){\makebox(0,0){\strut{}$s \; [M_\pi^2]$}}%
    }%
    \gplgaddtomacro\gplfronttext{%
      \csname LTb\endcsname%
      \put(5961,1683){\makebox(0,0)[r]{\strut{}E865 data}}%
      \csname LTb\endcsname%
      \put(5961,1463){\makebox(0,0)[r]{\strut{}NA48/2 data}}%
      \csname LTb\endcsname%
      \put(5961,1243){\makebox(0,0)[r]{\strut{}Dispersion relation, combined fit}}%
    }%
    \gplbacktext
    \put(0,0){\includegraphics{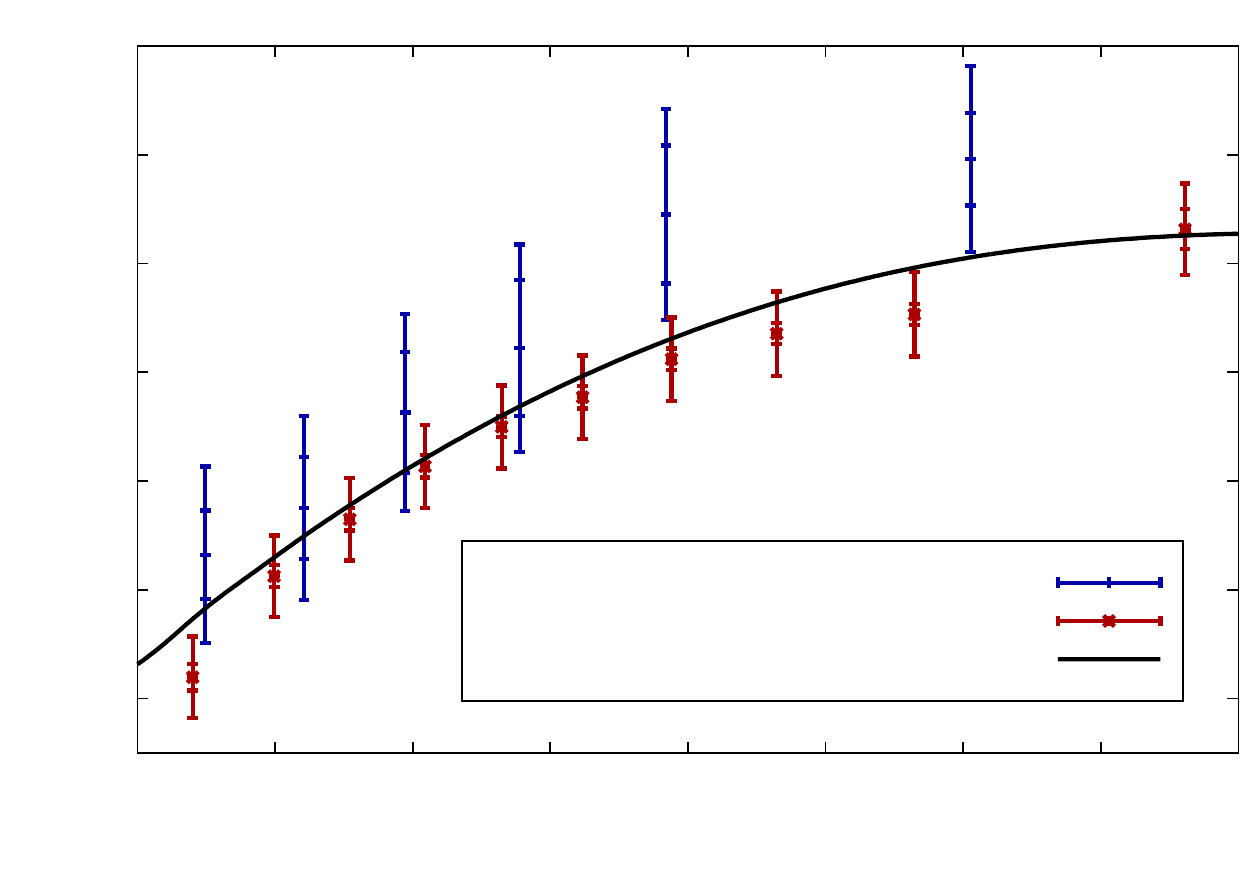}}%
    \gplfronttext
  \end{picture}%
\endgroup

%% file: plots/comparison_DR_data_pwave.tex
\begingroup
  \makeatletter
  \providecommand\color[2][]{%
    \GenericError{(gnuplot) \space\space\space\@spaces}{%
      Package color not loaded in conjunction with
      terminal option `colourtext'%
    }{See the gnuplot documentation for explanation.%
    }{Either use 'blacktext' in gnuplot or load the package
      color.sty in LaTeX.}%
    \renewcommand\color[2][]{}%
  }%
  \providecommand\includegraphics[2][]{%
    \GenericError{(gnuplot) \space\space\space\@spaces}{%
      Package graphicx or graphics not loaded%
    }{See the gnuplot documentation for explanation.%
    }{The gnuplot epslatex terminal needs graphicx.sty or graphics.sty.}%
    \renewcommand\includegraphics[2][]{}%
  }%
  \providecommand\rotatebox[2]{#2}%
  \@ifundefined{ifGPcolor}{%
    \newif\ifGPcolor
    \GPcolortrue
  }{}%
  \@ifundefined{ifGPblacktext}{%
    \newif\ifGPblacktext
    \GPblacktexttrue
  }{}%
  \let\gplgaddtomacro\g@addto@macro
  \gdef\gplbacktext{}%
  \gdef\gplfronttext{}%
  \makeatother
  \ifGPblacktext
    \def\colorrgb#1{}%
    \def\colorgray#1{}%
  \else
    \ifGPcolor
      \def\colorrgb#1{\color[rgb]{#1}}%
      \def\colorgray#1{\color[gray]{#1}}%
      \expandafter\def\csname LTw\endcsname{\color{white}}%
      \expandafter\def\csname LTb\endcsname{\color{black}}%
      \expandafter\def\csname LTa\endcsname{\color{black}}%
      \expandafter\def\csname LT0\endcsname{\color[rgb]{1,0,0}}%
      \expandafter\def\csname LT1\endcsname{\color[rgb]{0,1,0}}%
      \expandafter\def\csname LT2\endcsname{\color[rgb]{0,0,1}}%
      \expandafter\def\csname LT3\endcsname{\color[rgb]{1,0,1}}%
      \expandafter\def\csname LT4\endcsname{\color[rgb]{0,1,1}}%
      \expandafter\def\csname LT5\endcsname{\color[rgb]{1,1,0}}%
      \expandafter\def\csname LT6\endcsname{\color[rgb]{0,0,0}}%
      \expandafter\def\csname LT7\endcsname{\color[rgb]{1,0.3,0}}%
      \expandafter\def\csname LT8\endcsname{\color[rgb]{0.5,0.5,0.5}}%
    \else
      \def\colorrgb#1{\color{black}}%
      \def\colorgray#1{\color[gray]{#1}}%
      \expandafter\def\csname LTw\endcsname{\color{white}}%
      \expandafter\def\csname LTb\endcsname{\color{black}}%
      \expandafter\def\csname LTa\endcsname{\color{black}}%
      \expandafter\def\csname LT0\endcsname{\color{black}}%
      \expandafter\def\csname LT1\endcsname{\color{black}}%
      \expandafter\def\csname LT2\endcsname{\color{black}}%
      \expandafter\def\csname LT3\endcsname{\color{black}}%
      \expandafter\def\csname LT4\endcsname{\color{black}}%
      \expandafter\def\csname LT5\endcsname{\color{black}}%
      \expandafter\def\csname LT6\endcsname{\color{black}}%
      \expandafter\def\csname LT7\endcsname{\color{black}}%
      \expandafter\def\csname LT8\endcsname{\color{black}}%
    \fi
  \fi
  \setlength{\unitlength}{0.0500bp}%
  \begin{picture}(7200.00,5040.00)%
    \gplgaddtomacro\gplbacktext{%
      \csname LTb\endcsname%
      \put(660,704){\makebox(0,0)[r]{\strut{} 4}}%
      \put(660,1247){\makebox(0,0)[r]{\strut{} 4.2}}%
      \put(660,1790){\makebox(0,0)[r]{\strut{} 4.4}}%
      \put(660,2332){\makebox(0,0)[r]{\strut{} 4.6}}%
      \put(660,2875){\makebox(0,0)[r]{\strut{} 4.8}}%
      \put(660,3418){\makebox(0,0)[r]{\strut{} 5}}%
      \put(660,3961){\makebox(0,0)[r]{\strut{} 5.2}}%
      \put(660,4504){\makebox(0,0)[r]{\strut{} 5.4}}%
      \put(792,484){\makebox(0,0){\strut{} 4}}%
      \put(1585,484){\makebox(0,0){\strut{} 4.5}}%
      \put(2377,484){\makebox(0,0){\strut{} 5}}%
      \put(3170,484){\makebox(0,0){\strut{} 5.5}}%
      \put(3963,484){\makebox(0,0){\strut{} 6}}%
      \put(4755,484){\makebox(0,0){\strut{} 6.5}}%
      \put(5548,484){\makebox(0,0){\strut{} 7}}%
      \put(6340,484){\makebox(0,0){\strut{} 7.5}}%
      \put(7133,484){\makebox(0,0){\strut{} 8}}%
      \put(-110,2739){\rotatebox{-270}{\makebox(0,0){\strut{}$P$-wave: $|f_1/(PL \cdot \sigma_\pi)|$}}}%
      \put(3962,154){\makebox(0,0){\strut{}$s \; [M_\pi^2]$}}%
    }%
    \gplgaddtomacro\gplfronttext{%
      \csname LTb\endcsname%
      \put(5961,1683){\makebox(0,0)[r]{\strut{}E865 data}}%
      \csname LTb\endcsname%
      \put(5961,1463){\makebox(0,0)[r]{\strut{}NA48/2 data}}%
      \csname LTb\endcsname%
      \put(5961,1243){\makebox(0,0)[r]{\strut{}Dispersion relation, combined fit}}%
    }%
    \gplbacktext
    \put(0,0){\includegraphics{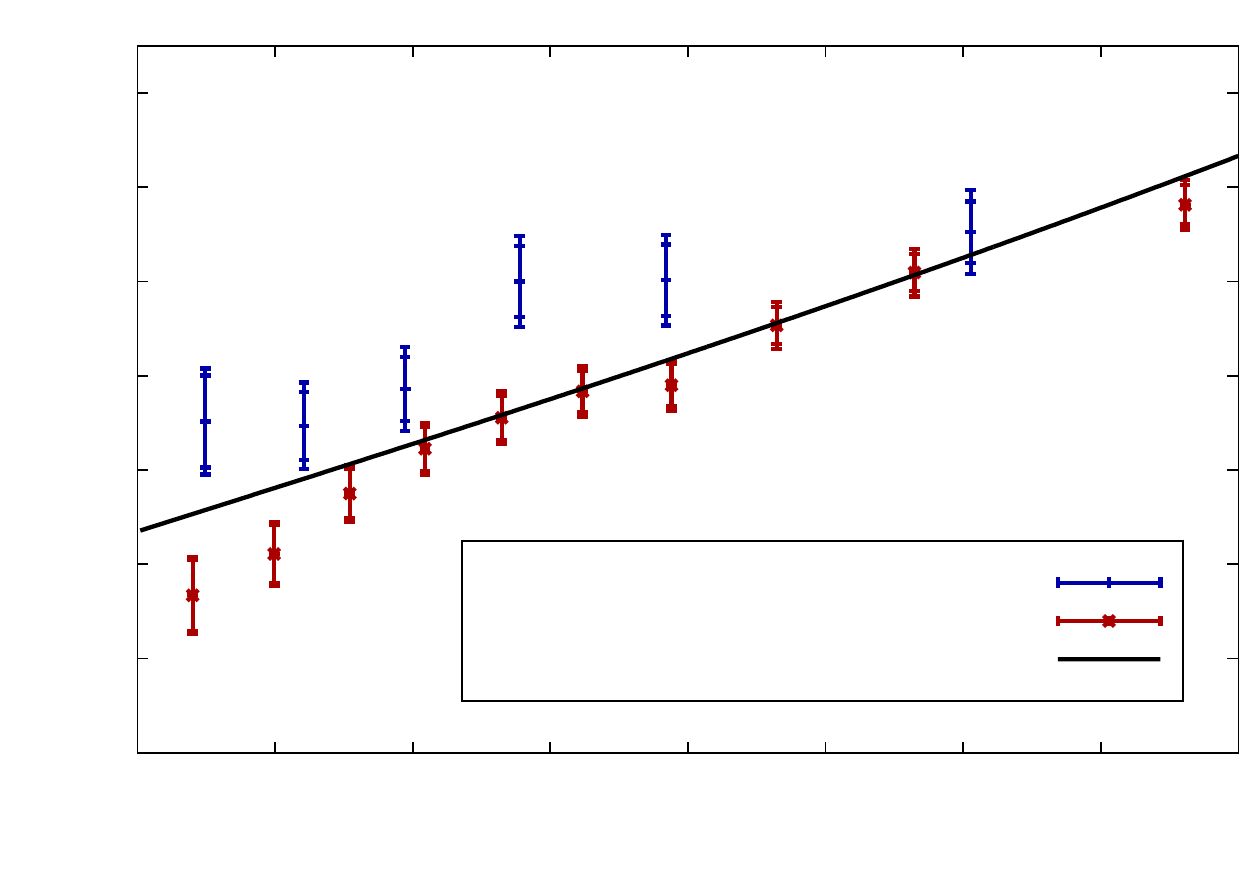}}%
    \gplfronttext
  \end{picture}%
\endgroup